\documentclass[paper]{JHEP3}

\usepackage{amsthm,graphicx}

\def\beq{\begin{equation}}
\def\eeq{\end{equation}}
\def\ber{\begin{eqnarray}}
\def\eer{\end{eqnarray}}

\def\m{{\rm m}}
\def\om{\Omega_{\rm m}}

\def\etal{{et al.}}

\def\tot{{\rm tot}}
\def\m{{\rm m}}

\title{Induced cosmological constant and other features of asymmetric brane embedding}

\author{Yuri Shtanov\\ Bogolyubov Institute for Theoretical Physics, Kiev 03680, Ukraine\\
E-mail: \email{shtanov@bitp.kiev.ua}}
\author{Varun Sahni\\ Inter-University Centre for Astronomy and Astrophysics, Post Bag 4,
Ganeshkhind, Pune 411~007, India\\ E-mail: \email{varun@iucaa.ernet.in}}
\author{Arman Shafieloo\\ Department of Physics, University of Oxford, 1 Keble Road, Oxford OX1~3RH,
UK\\ E-mail: \email{a.shafieloo1@physics.ox.ac.uk}}
\author{Alexey Toporensky\\ Sternberg Astronomical Institute, Moscow State University, Universitetsky
Prospekt, 13, Moscow 119992, Russia\\ E-mail: \email{lesha@xray.sai.msu.ru}}

\abstract{We investigate the cosmological properties of an `induced gravity' brane scenario in the
absence of mirror symmetry with respect to the brane. We find that brane evolution can
proceed along one of four distinct branches. By contrast, when mirror symmetry is
imposed, only two branches exist, one of which represents the self-accelerating brane,
while the other is the so-called normal branch. This model incorporates many of the
well-known possibilities of brane cosmology including phantom acceleration ($w < -1$),
self-acceleration, transient acceleration, quiescent singularities, and cosmic mimicry.
Significantly, the absence of mirror symmetry also provides an interesting way of
inducing a sufficiently small cosmological constant on the brane. A small (positive)
$\Lambda$-term in this case is induced by a small asymmetry in the values of bulk
fundamental constants on the two sides of the brane.}


\preprint{\arXivid{0901.3074}}

\begin{document}

\section{Introduction}

Extra-dimensional `braneworld' models have attracted considerable attention in recent
years. This is partly due to the fact that superstring/M-theory can only be consistently
formulated in a universe which has more than four space-time dimensions. A distinctive
feature of this theory is that it allows some of the extra dimensions to be large and
even infinite thereby accommodating the braneworld scenario \cite{hwitten}. It is now
well known that braneworld cosmologies can display quite distinctive behaviour which
departs from that in general relativity either during early or late times
\cite{brane_review}. The former can modify standard inflationary predictions for
primordial fluctuations while the latter can cause late-time acceleration. This latter
class of models will form the focus of our present paper. An abundance of recent
cosmological observations points to a universe which is currently accelerating
\cite{data}. Although the source of cosmic acceleration remains unknown, most
observations are well described by a small cosmological constant $\Lambda/8\pi G \simeq
10^{-47}\, \mbox{GeV}^4$. Such a small value for the $\Lambda$-term is difficult to
explain within the context of standard field theory, which typically predicts a value for
$\Lambda$ which is several orders of magnitude larger than what is indicated by
observations \cite{DE_review}.

In this setting it is natural to ask whether cosmic acceleration could arise via an
infrared modification of gravity at large distances. A famous example of such a scenario
is the  Dvali--Gabadadze--Porrati (DGP) model \cite{DGP}, which has been extensively
discussed in many papers \cite{brane_review}. The present work shall examine a family of
braneworld models which contain the DGP scenario as a subclass. As we shall show, this
family has many interesting features including that of phantom acceleration ($w_{\rm eff}
< -1$) and the possibility of inducing a small cosmological constant on the brane through
bulk effects.

Our braneworld scenario will contain a single large extra dimension. Two possibilities of
principle exist in this case: either the bulk space is constrained to be symmetric with
respect to the $Z_2$ group of reflections relative to the brane, or such a symmetry is
not imposed. The case where the bulk is symmetric is equivalent to the geometrical
setting in which the brane is just a boundary of the bulk space; this is one way of
justifying this mirror symmetry. An embedded brane without the $Z_2$ symmetry is,
however, a more general case with rich possibilities for cosmology. In spite of a
considerable number of papers on this class of braneworld models (see, e.g.,
\cite{asym,stealth,CGP,KPS}), some of these possibilities were either not noted or
insufficiently studied. They form the subject of the present work.

We consider a braneworld model described by the following simple yet generic action,
which includes
gravitational and cosmological constants in the bulk (${\mathcal B}$) and on the brane:
\begin{equation} \label{action}
S = \sum_{i = 1,2} M_i^3  \left[ \int_{{\mathcal B}_i} \left({\mathcal
R}_i - 2 \Lambda_i \right) - 2 \int_{\rm brane} K_i  \right] +
\int_{\rm brane} \left( m^2 R - 2 \sigma \right) + \int_{\rm brane} L
(h_{ab}, \phi) \, .
\end{equation}
Here, ${\mathcal R}_i$ is the scalar curvature of the five-dimensional metric $g^i_{ab}$
on ${\mathcal B}_i$, $i = 1, 2$, the two bulk spaces on either side of the brane, and $R$
is the scalar curvature of the induced metric $h_{ab}$ on the brane. The quantity $K_i =
K^i_{ab} h^{ab}$ is the trace of the symmetric tensor of extrinsic curvature $K^i_{ab}$
of the brane in the space ${\mathcal B}_i$. The symbol $L (h_{ab}, \phi)$ denotes the
Lagrangian density of the four-dimensional matter fields $\phi$ the dynamics of which is
restricted to the brane so that they interact only with the induced metric $h_{ab}$.  All
integrations over ${\mathcal B}_i$ and over the brane are taken with the corresponding
natural volume elements. The symbols $M_i$, $i = 1,2$, and $m$ denote the Planck masses
of the corresponding spaces, $\Lambda_i$, $i = 1,2$, are the five-dimensional
cosmological constants on the two side of the brane, and $\sigma$ is the brane tension.
We shall focus on the asymmetric case with $\Lambda_1 \neq \Lambda_2$ and $M_1 \neq M_2$
which appears to be preferable from a string-theory perspective. For instance, the
dilaton stabilised  in different vacuum states on adjacent sides of the brane would lead
to an effective five-dimensional theory with $M_1 \neq M_2$. The string landscape is
likely to favour $\Lambda_1 \neq \Lambda_2$, which also occurs in domain wall scenarios.
(The Randall--Sundrum (RS) \cite{RS} and Dvali--Gabadadze--Porrati (DGP) \cite{DGP}
models can be derived from (\ref{action}) when mirror symmetry is respected.)

In the absence of $Z_2$ symmetry, cosmological evolution of the brane is described by
\begin{equation} \label{cosmol}
H^2 + {\kappa \over a^2} = {\rho + \sigma \over 3 m^2} + {1 \over m^2}
\sum_{i = 1,2} \zeta^{}_i M_i^3 \sqrt{H^2 + {\kappa \over a^2} - {\Lambda_i \over 6} -
{C_i \over a^4} } \, ,
\end{equation}
where $\rho$ is the total energy density of matter on the brane, $\zeta_i = \pm 1$, $i =
1,2$, correspond to the two possible ways of bounding each of the bulk spaces ${\mathcal
B}_i$, $i = 1,2$, by the brane. We classify the resulting four branches according to the
signs of $\zeta_1$ and $\zeta_2$ as $(++)$, $(+-)$, $(-+)$, or $(--)$.\footnote{Note
that, in the case of $Z_2$ symmetry, there are only two ways of bounding the bulk by the
brane, and these were called brane\,1 and brane\,2 in \cite{ss02}. Of these, brane\,2
contains the self-accelerating DGP brane as a subclass, while brane\,1 can lead to
phantom acceleration.}

In the limit of $Z_2$ symmetry, the branches $(--)$ and $(++)$ become the normal branch
(brane\,1) and self-accelerating branch (brane\,2), respectively.  The other two
so-called {\em mixed\/} branches are characterised by $\zeta_1 \zeta_2 = -1$. As we shall
show in this paper, cosmology on these branches can be quite novel. For instance, a small
asymmetry in the values of the bulk constants on adjacent sides of the brane can induce a
cosmological constant on the brane itself. On the other hand, in the limit when the bulk
constants become exactly equal, one obtains stealth branes \cite{stealth} that obey the
general-relativistic internal equations but do not affect the metric of the bulk space.

In this paper, we study the implications of (\ref{cosmol}) for a spatially flat universe
($\kappa = 0$) without dark radiation ($C_i = 0$, $i = 1,2$). Equation (\ref{cosmol})
then simplifies to
\begin{equation} \label{main}
H^2  = {\rho + \sigma \over 3 m^2} + {1 \over m^2} \sum_{i = 1,2} \zeta^{}_i M_i^3
\sqrt{H^2 - {\Lambda_i \over 6} }
= {\rho + \sigma \over 3 m^2} + \sum_{i = 1,2} {\zeta_i \over \ell_i}
\sqrt{H^2 + \lambda_i^{-2} } \, ,
\end{equation}
where we have introduced the fundamental lengths
\begin{equation} \label{lengths}
\ell_i = {m^2 \over M_i^3} \, , \qquad \lambda_i = \sqrt{ - {6 \over \Lambda_i}} \, ,
\qquad i = 1,2 \, ,
\end{equation}
assuming negative values of the bulk cosmological constants.

Note that (\ref{main}) can be rewritten in terms of an {\em effective} cosmological constant,
$\Lambda_{\rm eff}$, as
\beq \label{LCDM}
H^2 = \frac{\rho}{3m^2} + {\Lambda_{\rm eff} \over 3} \, ,
\eeq
where
\beq
{\Lambda_{\rm eff} \over 3} = {\sigma \over 3 m^2} + \sum_{i = 1,2} {\zeta_i \over \ell_i}
\sqrt{H^2 + \lambda_i^{-2} }~,
\label{lam}
\eeq
which will be useful to us when we study the cosmological properties of this
braneworld later in this paper.
A pictorial representation of the branches described by (\ref{main}) is given in the
appendix.

In this paper, we consider the late-time evolution of the universe, in which the energy
density $\rho$ is dominated by matter with the equation of state $p = 0$.  Then,
introducing the cosmological parameters
\begin{equation} \label{omegas}
\om = {\rho_0 \over 3 m^2 H_0^2} \, , \qquad \Omega_\sigma = {\sigma \over 3 m^2 H_0^2} \, ,
\qquad \Omega_{\ell_i} = \ell_i^{-2} H_0^{-2} \, , \qquad
\Omega_{\lambda_i} = \lambda_i^{-2} H_0^{-2} \, ,
\end{equation}
where $\rho_0$ and $H_0$ are the current values of the matter density and Hubble
parameter, respectively, we rewrite (\ref{main}) in terms of the cosmological redshift
$z\,$:
\begin{equation} \label{redshift}
h^2 (z) \equiv {H^2 (z) \over H_0^2} = \om (1 + z)^3 + \Omega_\sigma + \sum_{i = 1,2} \zeta_i
\sqrt{\Omega_{\ell_i}} \sqrt{h^2 (z) + \Omega_{\lambda_i}} \, .
\end{equation}
This equation implicitly determines the function $h(z)$, and explicitly the inverse
function $z (h)$. Note that the dimensionless cosmological parameters are related through
the constraint equation
\beq
\om+\Omega_\sigma + \sum_{i = 1,2}\zeta_i\sqrt{\Omega_{\ell_i}}\sqrt{1+\Omega_{\lambda_i}} = 1~.
\label{constraint}
\eeq

We now proceed to describe some features of braneworld
cosmology without $Z_2$ symmetry.

\section{Induced cosmological constant on the brane}

One way of accounting for cosmic acceleration within the framework of braneworld theory
with mirror symmetry was suggested by Dvali, Gabadadze and Porrati \cite{DGP}. An
extension of this model to the case when mirror symmetry is absent is obtained by setting
to zero the cosmological constants on the brane and in the bulk, so that $\sigma = 0$,
$\Lambda_i = 0$, $i = 1,2$.  The expansion law (\ref{main}) then simplifies to
\beq \label{dgp0}
H^2 - H\sum_{i = 1,2}\frac{\zeta_i}{\ell_i} = \frac{\rho}{3m^2} \, ,
\eeq
which evolves to a De~Sitter limit at late times
\begin{equation} \label{dgp}
\lim_{z \to -1} H (z) = H_{\rm DS} = \sum_{i = 1,2}\frac{\zeta_i}{\ell_i} \, ,
\end{equation}
provided $\sum\limits_{i=1,2}\zeta_i/\ell_i$ is positive, which is true for branches
$(++)$ and $(+-)$, provided $M_1 > M_2$ in the latter case.  If $m$ is of the order of
the planck mass $M_{\rm P} \simeq 10^{19}\,\mbox{GeV}$, then the values of $M_i \sim
100\,\mbox{MeV}$ can explain the observed cosmic acceleration. [The self-accelerating DGP
model corresponds to the $(++)$ branch with $\ell_1 = \ell_2$.]

The absence of mirror symmetry provides a new avenue for this mechanism. Specifically, the
observed cosmic acceleration can be produced on one of the mixed branches with {\em
arbitrarily high\/} values of the bulk Planck masses $M_1$ and $M_2$, provided these
values are sufficiently close to each other. If $0 < \Delta M \equiv M_1 - M_2 \ll M_1$,
then we have
\beq \label{hds0}
H_{\rm DS} = \frac{M_1^3 - M_2^3}{m^2} \approx \frac{3 M_1^2}{m^2} \Delta M
\eeq
on the $(+-)$ branch, and, by adjusting the value of $\Delta M$, one can always achieve
an observationally suitable value of $H_{\rm DS}$.  For example, if $M_1, M_2 \sim m$,
then one needs $\Delta M \sim H_0$.

The previous model gave one example of late-time acceleration in the absence of the
(brane) cosmological constant. We now derive another model with the same property but
with a more flexible assumption $\Lambda_i \ne 0$. Setting $\sigma = 0$ and $\Lambda_i
\ne 0$ in (\ref{main}) leads to
\beq
H^2 - \sum_{i = 1,2}\frac{\zeta_i}{\ell_i}\sqrt{H^2-{\Lambda_i \over 6}}
 = \frac{\rho}{3m^2} ~,
\label{newcosmos}
\eeq
which evolves to a different De~Sitter limit, expressed by the equation
\begin{equation} \label{desitter}
\lim_{z \to -1} H^2 (z) = H_{\rm DS}^2  = \sum_{i = 1,2} \frac{\zeta_i}{\ell_i}\sqrt{H_{\rm DS}^2
+ \lambda_i^{-2} } \, ,
\end{equation}
where the length scales $\ell_i$ and $\lambda_i$ are defined in (\ref{lengths}).

It is interesting that a tiny asymmetry between the two bulk spaces can lead to a small
cosmological constant being induced on the brane.  Provided the bulk parameters $M_1$ and
$M_2$ as well as $\Lambda_1$ and $\Lambda_2$ are close to each other, a neat cancelation
on the right-hand side of (\ref{desitter}), which occurs for $\zeta_1 \zeta_2 = -1$,
leads to a small value of $H_{\rm DS}$. Remarkably, this can happen even for very large
values of the bulk constants. In particular, assuming that $\lambda_i \ll H_{\rm
DS}^{-1}$, we have
\begin{equation} \label{hds}
H_{\rm DS}^2  \approx  \left| {1 \over \ell_1 \lambda_1} - {1 \over \ell_2 \lambda_2} \right|
\end{equation}
for one of the mixed branches.  Thus, for bulk parameters of the order of a TeV, $M_i
\sim 1\,\mbox{TeV}$, $\lambda_i \sim 1\,\mbox{TeV}^{-1}$, we recover the current value of
the Hubble parameter ($H_{\rm DS} \sim H_0$) provided
\begin{equation} \label{lambda0}
\left| \ell_1 \lambda_1 - \ell_2 \lambda_2 \right|^{1/2} \sim 10^{-13}\,\mbox{TeV}^{-1}
\sim 10^{-30}\,\mbox{cm} \, .
\end{equation}

Equations such as (\ref{hds0}) or (\ref{hds}), (\ref{lambda0}) certainly represent fine
tuning, with a tiny difference between bulk parameters only slightly breaking the
smoothness of the metric across the brane.\footnote{Perhaps, the small asymmetry in the
fundamental constants characterising the bulk can be explained by the presence of the
brane itself. For instance, the presence of the brane could lead to a small difference in
the quantum contribution to the effective action of the bulk on its two sides, inducing
slightly different bulk constants.} In the limit of exact equality of the bulk constants
on the two sides of the brane, the branches with $\zeta_1 \zeta_2 = -1$ describe a smooth
bulk space, and the brane approaches the limit of a stealth brane \cite{stealth},
evolving according to the usual Einstein equations without affecting the bulk space.

As we shall see in the next section, the cosmological scenario with an induced
cosmological constant is distinguished by a property called {\em cosmic mimicry\/} which
has interesting observational signatures.

\section{Braneworld expansion can mimic $\Lambda$CDM}

For large values of the bulk parameters, we encounter the phenomenon of {\em cosmic
mimicry\/} which, in the context of $Z_2$ symmetry, was described in \cite{mimicry}. Note
that, during the radiation and matter-dominated epochs, the expansion of the universe
follows the general-relativistic prescription
\begin{equation}
H^2 \approx {\rho + \sigma \over 3 m^2} \, ,
\end{equation}
where $\sigma/m^2$ plays the role of the cosmological constant on the brane. However, at
very late times, cosmic expansion gets modified due to extra-dimensional effects. Indeed,
if $\lambda_i \ll H_0^{-1}$, then the square root in the last term of (\ref{main}) can be
expanded in the small parameter $\lambda_i^2 H^2$ at late times, and the braneworld
expands according to $\Lambda$CDM, namely
\beq
H^2 = \frac{8\pi G\rho}{3} + {\Lambda \over 3} \label{LCDM1}
\eeq
with
\ber
8 \pi G &=& m^{-2} \left(1 - \sum_{i = 1,2} {\zeta_i \lambda_i \over 2 \ell_i}
\right)^{-1} \, , \label{8piG}\\
\Lambda &=& \left( {\sigma \over m^2} + \sum_{i = 1,2} {3 \zeta_i \over \ell_i \lambda_i}
\right) \left(1 - \sum_{i = 1,2} {\zeta_i \lambda_i \over 2 \ell_i} \right)^{-1} \, .
\label{mimic2}
\eer

Note that both $G$ and $\Lambda$ are {\em independent of time\/}. Equations
(\ref{LCDM1})--(\ref{mimic2}) have important ramifications. They inform us that the
`bare' value of the cosmological constant on the brane, $\sigma$, is  `screened' at late
times by extra-dimensional effects resulting in its effective value $\Lambda$. Thus, the
early-time and late-time values of the cosmological constant are likely to be different,
and this makes our model open to verification.

Also note that one can have $\Lambda \neq 0$ even if $\sigma=0$.  Then, a small
$\Lambda$-term can be induced during late-time evolution on the brane {\em solely by
extra-dimensional effects\/}, as pointed out in the previous section.  The mechanism by
which the induced $\Lambda$-term becomes relatively small consists in a compensation of
two potentially large terms with opposite signs in equation (\ref{mimic2}). Specifically,
for small values of $\lambda_i$ (which correspond to large values of $\Lambda_i$) such
that $\lambda_i / \ell_i \ll 1$, in the case $\sigma = 0$, we have, approximately,
\beq
\Lambda \approx \sum_{i = 1,2} {3 \zeta_i \over \ell_i \lambda_i}  \, ,
\eeq
which is another form of the result (\ref{hds}) for one of the mixed branches.  What is
remarkable here is that a {\em positive\/} cosmological constant on the brane can be
sourced by bulk cosmological constants which are {\em negative\/}.

From (\ref{LCDM1}), (\ref{8piG}) we also find that the effective gravitational constants
during the early and late epochs are related by a multiplicative factor
\begin{equation}
1 - \sum_{i = 1,2} {\zeta_i \lambda_i \over 2 \ell_i}  \, ,
\end{equation}
which can be larger as well as smaller than unity, depending on the braneworld branch.
This factor will be closer to unity for the mixed branches ($\zeta_1 \zeta_2 = -1$) than
it is for the usual branches ($\zeta_1 \zeta_2 = 1$) which survive in the case of $Z_2$
symmetry.

Focussing on the important case where $\sigma=0$ and the effective four-dimensional
cosmological constant is induced entirely by five-dimensional effects, we find that, at
redshifts significantly below the {\em mimicry redshift\/}
\beq
z_\m \simeq \left (\frac{\Omega_{\lambda_i}}{\om}\right )^{1/3} - 1 \, , \qquad
\Omega_{\lambda_1} \simeq \Omega_{\lambda_2} \, ,
\eeq
the brane expansion mimics $\Lambda$CDM
\beq
h^2(z) = \widetilde\Omega_\m(1+z)^3 + \Omega_{\Lambda} \, , \qquad  z \ll z_\m \, ,
\eeq
with `screened' values of the
cosmological parameters:
\ber
\widetilde\Omega_\m &=& \om\left (1-\sum_{i = 1,2}\frac{\zeta_i}{2}
\sqrt{\frac{\Omega_{\ell_i}}{\Omega_{\lambda_i}}}\right )^{-1} \, ,  \\
\Omega_{\Lambda} &=& \sum_{i = 1,2}\zeta_i\sqrt{\Omega_{\ell_i}}
\sqrt{\Omega_{\lambda_i}} \left ({1 - \sum_{i =
1,2}\frac{\zeta_i}{2}\sqrt{\frac{\Omega_{\ell_i}}{\Omega_{\lambda_i}}}}\right )^{-1} \, .
\label{mimicry3}
\eer
On the other hand, from (\ref{LCDM1}) it follows that, at high redshifts, the universe
expands as SCDM
\beq
h^2(z) = \om(1+z)^3 \, , \qquad z\gg z_\m \, .
\eeq
An important distinguishing feature of this model is that the (screened) matter density,
$\widetilde\Omega_\m$, inferred via geometrical tests based on standard candles and
rulers, may not match its (bare) dynamical value $\om$. This allows cosmic mimicry to be
distinguished from other cosmological scenarios by means of the {\em Om diagnostic\/}
suggested in \cite{om}. The fact that brane expansion also follows different laws at low
and high redshift provides another important observational test of this model.

\section{Phantom branes}

In the presence of $Z_2$ symmetry, the brane\,1 branch of the generic model
(\ref{cosmol}) exhibits phantom-like behaviour \cite{ss02} which is in excellent
agreement with observations \cite{brane_obs} (see also \cite{ls04}). Let us see whether
this behaviour persists when mirror symmetry is absent. Note first that the condition for
phantom acceleration, $w (z) < -1$, where
\beq \label{w}
w(z) = {2 q(z) - 1 \over 3 \left[1 - \Omega_\m(z) \right] } \, , \quad q (z) =
\frac{d\log{H (z)}}{d \log (1+z)} - 1 \, , \quad \Omega_\m (z) =
\frac{\Omega_\m (1 + z)^3}{h^2(z)} \, ,
\eeq
has two equivalent formulations:
\beq
\Omega_\m(z) > \frac{2}{3} \frac{d\log{H (z)}}{d \log (1+z)} \quad \mbox{and} \quad {\dot
\Lambda_{\rm eff}} > 0 \, ,
\eeq
where $\Lambda_{\rm eff}$ is the effective cosmological constant in (\ref{lam}), and
differentiation is carried out with respect to the physical time variable. In the case of
the $(--)$ brane ($\zeta_1 = \zeta_2 = -1$), one has
\beq
\Lambda_{\rm eff} = {\sigma \over 3 m^2} - \sum_{i = 1,2} \frac{\sqrt{H^2 +
\lambda_i^{-2}}}{\ell_i} \, ,
\eeq
and we find immediately that $\Lambda_{\rm eff}$ {\em increases with time\/} when the
expansion rate, $H$, decreases.  It is also quite clear that one (and only one) of the
mixed branches will necessarily have a negative value of the sum term in (\ref{lam}),
again exhibiting phantom behaviour.

It is straightforward to verify that ${\dot \Lambda_{\rm eff}} > 0$ and ${\dot H} < 0$ on
the two branches exhibiting phantom behaviour. Differentiating (\ref{main}) and
(\ref{lam}), we find
\ber
{\dot H} &=& - \frac{\rho}{m^2}\left\lbrack 2 - \sum_{i =
1,2}\frac{\zeta_i}{\ell_i\sqrt{H^2+\lambda_i^{-2}}}\right\rbrack^{-1} \ < \ 0 \, , \\
{\dot \Lambda_{\rm eff}} &=& 3H{\dot H}\, \sum_{i = 1,2}
\frac{\zeta_i}{\ell_i\sqrt{H^2+\lambda_i^{-2}}} \ > \ 0 \\
\nonumber &{}& \mbox{for} \quad \sum_{i = 1,2}
\frac{\zeta_i}{\ell_i\sqrt{H^2+\lambda_i^{-2}}} < 0 \, .
\eer
Note that phantom models \cite{caldwell02} with constant equation of state, $w<-1$,
 are marked by ${\dot
\Lambda_{\rm eff}} > 0$ and {\em super-acceleration}: ${\dot H} > 0$ at late
times.\footnote{It is easy to show that, in phantom models, the turning point ${\dot
H}=0$ occurs at
\beq
1+z_* \equiv \frac{a_0}{a(t_*)}=\left (\frac{1-\om}{\om}|1+w|\right )^{1/3|w|} \, , \quad
w < - 1 \, .
\eeq}
This is related to the fact that the dark-energy (phantom) density in such models {\em
increases\/}, as the universe expands, according to
\beq
\rho_{\rm phantom} \propto a^{3|1+w|} \, , \quad  w < - 1 \, ,
\eeq
which causes the Hubble parameter to grow at late times,  eventually leading to a
Big-Rip singularity at which $H$ diverges. By contrast, although the
behaviour of our braneworld is phantom-like ($w_{\rm eff} < -1$), the universe never
super-accelerates since ${\dot H} < 0$ always holds. Furthermore, since $H$ decreases
during expansion, a Big-Rip-type future singularity
 which plagues phantom cosmology is
absent in the braneworld. From the definition
\beq
q = -\left (1+\frac{\dot H}{H^2}\right )
\eeq
and property ${\dot H} < 0$, we find $q>-1$. In fact, the deceleration parameter in our
model always remains larger than the De~Sitter value of $q = - 1$, approaching it only in
the limit of $t\to \infty$.

\section{Transient acceleration}

An important property of this class of braneworld models is that the current acceleration
of the universe need not be eternal. In other words, for a specific relationship between
the fundamental parameters in (\ref{main}), the acceleration of the universe is a {\em
transient\/} phenomenon, and the universe reverts back to matter-dominated expansion in
the future. Within the context of mirror symmetry, this scenario was called {\em
disappearing dark energy\/} and discussed in \cite{ss02}. In the absence of mirror
symmetry, it was studied in \cite{CGP} under the name {\em
stealth-acceleration\/} (which should not be confused with the `stealth brane' of
\cite{stealth}).

Transient acceleration implies the property $H \to 0$ in the asymptotic future, which
requires the following condition to be satisfied:
\beq\label{dde}
{\sigma \over 3 m^2} + \sum_{i = 1,2} {\zeta_i \over \ell_i \lambda_i} = 0 \quad
\Rightarrow \quad \Omega_\sigma + \sum_{i = 1,2}\zeta_i
\sqrt{\Omega_{\ell_i}\Omega_{\lambda_i}} = 0 \, .
\eeq
On the $(--)$ and $(++)$ branches, this condition is realised with the following
respective values of the brane tension:
\beq
{\sigma \over 3 m^2} = \pm \left( {1 \over \ell_1 \lambda_1} + {1 \over \ell_2 \lambda_2}
\right) \quad \Rightarrow \quad \Omega_\sigma = \pm  \left(
\sqrt{\Omega_{\ell_1}\Omega_{\lambda_1}} + \sqrt{\Omega_{\ell_2}\Omega_{\lambda_2}}\right
) \, . \label{transient1}
\eeq
On the new mixed branches $(+-)$ and $(-+)$, the required brane tension is smaller by
absolute value:
\begin{equation}\label{transient2}
{\sigma \over 3 m^2} = \pm \left| {1 \over \ell_1 \lambda_1} - {1 \over \ell_2 \lambda_2}
\right| \quad \Rightarrow \quad \Omega_\sigma = \pm \left| \sqrt{\Omega_{\ell_1}\Omega_{\lambda_1}} -
 \sqrt{\Omega_{\ell_2}\Omega_{\lambda_2}}\right| \, .
\end{equation}

Under constraint (\ref{dde}), the cosmological evolution equation (\ref{redshift})
becomes
\begin{equation} \label{dde-evolution}
h^2 (z) = \om (1 + z)^3 + \sum_{i = 1,2} \zeta_i \sqrt{\Omega_{\ell_i}}
\left( \sqrt{h^2 (z) + \Omega_{\lambda_i}} - \sqrt{\Omega_{\lambda_i}} \right) \, .
\end{equation}

Condition (\ref{dde}) is necessary but not sufficient to speak about transient
acceleration on a particular branch.  A distinguishing property of a transiently
accelerating brane is that $q(z) \to 0.5$ in the remote past ($10^5 \gg z \gg 1$) as well
as in the remote future ($z \to -1$), reflecting the fact that the universe is matter
dominated in the past and in the future, while, during the current phase, the
deceleration parameter is negative, $q_0 < 0$.  This last condition is realised only if
the cosmological expansion law $H^2 ( \rho)$ is convex upwards to a sufficiently high
degree. Specifically, in view of the second expression in (\ref{w}), the condition $q_0 <
0$ can be presented in the form
\begin{equation}
{d H^2 (\rho_0) \over d \rho} < { 2 H_0^2 \over 3\rho_0} \, .
\end{equation}
Looking at figures \ref{fig:intersect} and \ref{fig:nonintersect} in the appendix, one
can see that this property can be realised only on two of the four branches: on the
$(++)$ branch and on one of the mixed branches.

The expression for the current value of the deceleration parameter can be calculated by
using the formula
\begin{equation}
q_0 =  { 3 \Omega_\m \over 2 - \sum_i \zeta_i \sqrt{\displaystyle \Omega_{\ell_i}
\over \displaystyle  1 + \Omega_{\lambda_i}} } - 1 \, .
\end{equation}
One should note that only four out of five $\Omega$ parameters are independent in  this
expression because of the normalization condition $h^2(0) = 1$ applied to the evolution
equation (\ref{dde-evolution}).  In the $Z_2$-symmetric case, there remain only two
independent $\Omega$ parameters.  It is clear then that transient acceleration can be
realised more easily in the $Z_2$-asymmetric case.  This is illustrated in figures
\ref{fig:symmetric} and \ref{fig:asymmetric}, which show the corresponding behaviour of
the deceleration parameter $q(z)$.

\FIGURE{\label{fig:symmetric}
\includegraphics[width=0.47\textwidth,angle=-90]{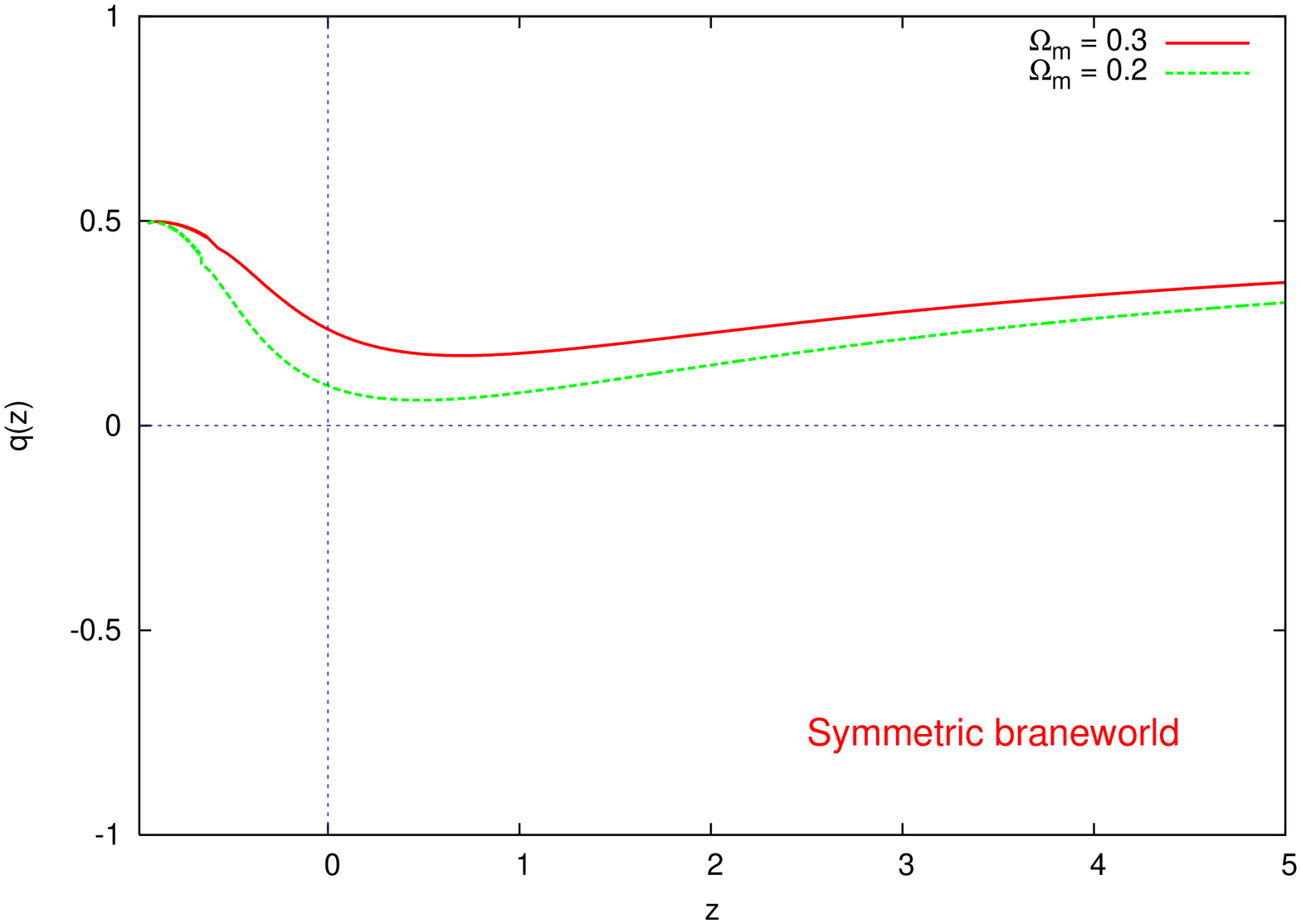}
\caption{The deceleration parameter versus redshift is plotted for the $(++)$ branch in
the case of $Z_2$ symmetry. The model has the parameters $\Omega_{\lambda_1} =
\Omega_{\lambda_2} = 2$.  We present plots for two different values of the matter density
parameter: $\om = 0.3$ and $\om = 0.2$. The sets of other parameters are calculated to be
$\Omega_{\ell_1} = \Omega_{\ell_2} = 1.21$, $\Omega_\sigma = - 3.11$ and $\Omega_{\ell_1}
= \Omega_{\ell_2} = 1.58$, $\Omega_\sigma = - 3.56$, respectively. In this case, an
accelerated regime is not realised, although deceleration is significantly slowed down at
the present cosmological epoch.}}

\FIGURE{\label{fig:asymmetric}
\includegraphics[width=0.47\textwidth,angle=-90]{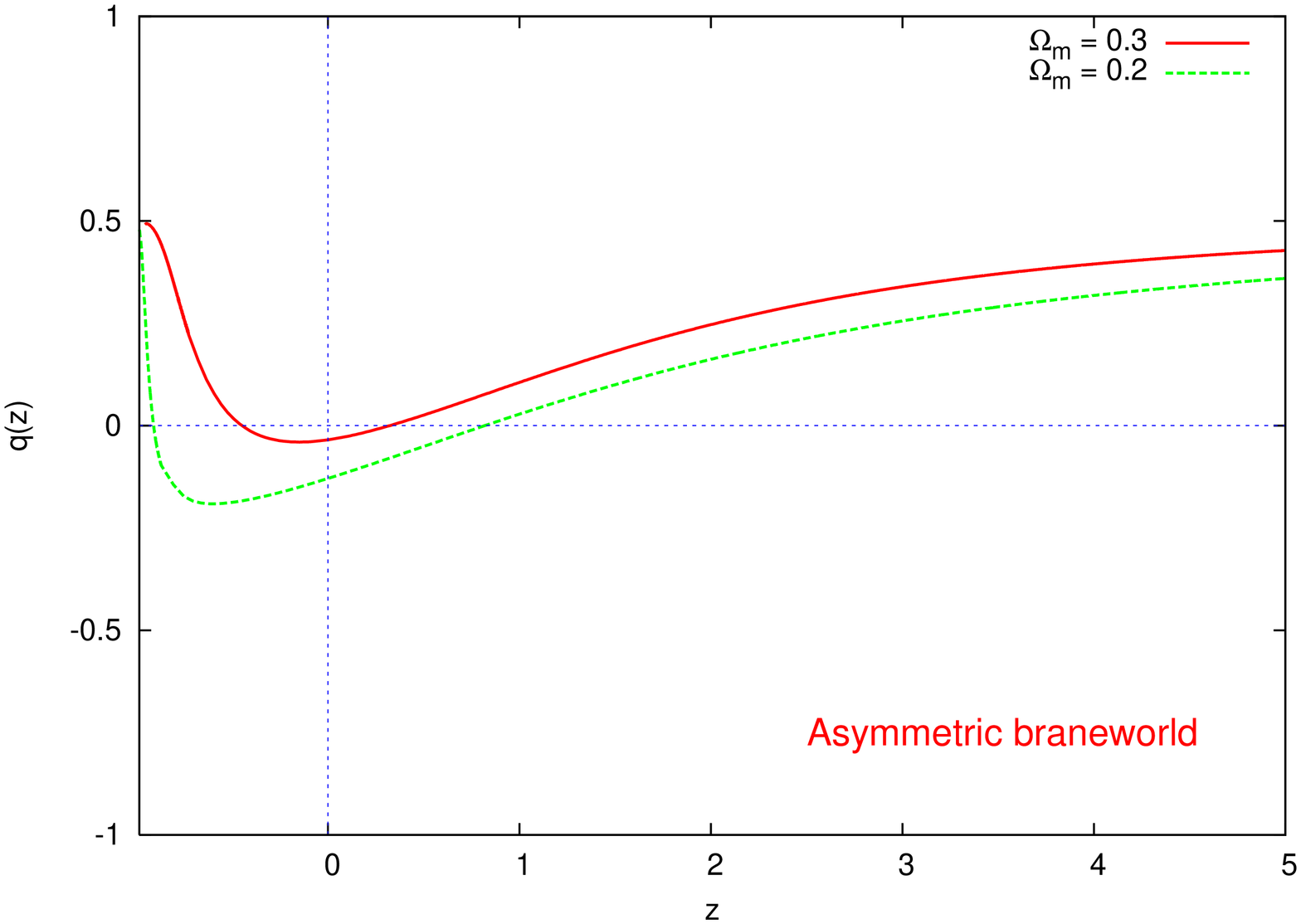}
\caption{The deceleration parameter versus redshift is plotted for the $(+-)$ branch in
the case of absence of $Z_2$ symmetry. The model has the parameters $\Omega_{\lambda_1} =
2$ and  $\Omega_{\lambda_2} = 2.1$.  We present plots for two different sets of values of
the remaining two independent parameters: $\left( \om, \Omega_{\ell_1} \right) = (0.3,\,
10000)$, which results in $\left(\Omega_{\ell_2}, \Omega_\sigma \right) = (9954.68,\,
3.16)$, and $\left( \om, \Omega_{\ell_1} \right) = (0.2,\, 5000)$, which results in
$\left(\Omega_{\ell_2}, \Omega_\sigma \right) = (4840.15,\, 0.82)$. Both plots show
acceleration at the present cosmological epoch, which generically becomes more prominent
for lower values of $\om$.}}

In a transiently accelerating universe, cosmic acceleration is sandwiched between two
matter-dominated regimes. A transiently accelerating braneworld clearly does not possess
the Big Rip of phantom cosmology, nor even the event horizon of De~Sitter space\,! An
in-depth study of this class of models \cite{CGP} has revealed the existence of regions
in parameter space which are stable (ghost-free).

We have demonstrated that
it is possible to construct braneworld models with transient acceleration.
What is less clear is
whether such transiently accelerating branches will pass key cosmological
tests based on observations of high-redshift type Ia supernovae, baryon acoustic
oscillations, etc. This important issue is open for further study.

\section{Quiescent singularities}

A new feature of brane cosmology is a possible presence of {\em quiescent\/}
singularities at which the density, pressure and expansion rate remain finite, while the
deceleration parameter and the Kretchman invariant, $R_{iklm}R^{iklm}$, diverge
\cite{quiescent}.  The universe encounters such a singularity in the future if a point is
reached during expansion where the derivative of $H^2$ with respect to $\rho$ goes to
infinity or, equivalently, where the derivative of $\rho$ with respect to $H^2$ vanishes.
Using (\ref{main}), we can express this condition as the existence of a positive root
$H_s^2$ of the equation
\begin{equation} \label{sing}
\sum_{i = 1,2} {\zeta_i \over \ell_i \sqrt{H_s^2 + \lambda_i^{-2} } } = 2 \, ,
\end{equation}
and a quiescent singularity is approached as $H \to H_s$. At this moment, expansion
formally ceases, and one cannot extend the classical evolution of the brane beyond this
point.
  Such a singular point
obviously exists on the $(++)$ branch if and only if
\begin{equation}
{\lambda_1 \over \ell_1} + {\lambda_2 \over \ell_2} > 2 \quad
\Rightarrow \quad \sqrt{\frac{\Omega_{\ell_1}}{\Omega_{\lambda_1}}}
+ \sqrt{\frac{\Omega_{\ell_2}}{\Omega_{\lambda_2}}} > 2 \, ,
\end{equation}
and it is reachable on this branch if the brane tension $\sigma$ is sufficiently
negative:
\begin{equation}
{\sigma \over 3 m^2} < H_s^2 - \sum_{i = 1,2} {1 \over \ell_i} \sqrt{ H_s^2 +
\lambda_i^{-2}} < 0 \, ,
\end{equation}
or, equivalently,
\beq
\Omega_\sigma < \frac{H_s^2}{H_0^2} - \sum_{i = 1,2}
\sqrt{\Omega_{\ell_i}}\sqrt{\frac{H_s^2}{H_0^2} + \Omega_{\lambda_i}} < 0 \, .
\eeq

Condition (\ref{sing}) may or may not be realised on the mixed branches. For example, in
the simplifying case $\ell_1 = \ell_2 = \ell$, condition (\ref{sing}) is realised on the
mixed branch $(+-)$ provided $\lambda_1 > \lambda_2$.  One can show that the values of
the parameters $\ell_i$, $\lambda_i$, $i = 1,2$, in principle can be chosen so that
equation (\ref{sing}) has positive roots for three branches $(++)$, $(+-)$ and $(-+)$. To
achieve this, one only needs to satisfy the conditions $\ell_1 > \ell_2$ and $\lambda_1
\ell_2 > \lambda_2 \ell_1$ and choose sufficiently small values of $\ell_1$, $\ell_2$.

For a graphical presentation of the reasons for the existence of quiescent singularities,
the reader can look into the appendix. As in the case of mirror symmetry, quantum effects
may play an important role in the vicinity of a quiescent singularity \cite{quantum}; see
also \cite{sing}.

We also note that, in the case of mirror symmetry, realisation of quiescent singularity
requires either negative brane tension or positive bulk cosmological constant (both
conditions are suspicious from the viewpoint of possible instabilities). However the
quiescent singularity can easily be realised without these assumptions in the asymmetric
case on a mixed branch.

The presence of a quiescent singularity in the future of the cosmological evolution does
not threaten the past cosmological scenario.  Therefore, this issue, just like the issue
of Big Rip of phantom cosmology, is mainly of academic interest.  Here, we only wish to
point out that the possibility of quiescent singularity can be realised rather easily in
braneworld theory in certain domain of its parameters without any additional ingredients
(such as phantom fields, which lead to Big Rip singularities).

\section{Discussion}

In this paper, we derive the expansion laws for an induced-gravity brane model in the
general case where mirror ($Z_2$) symmetry of the bulk space with respect to the brane
may or may not be present. We find that, depending upon the choice of brane embedding,
cosmological expansion on the brane can proceed along four independent branches, two of
which survive in the case of $Z_2$ symmetry. An important property of this class of
models is that the four-dimensional gravitational and cosmological constants are
effective quantities derivable from five-dimensional physics. In this case, brane
expansion mimics $\Lambda$CDM at low redshifts, but the `screened' matter density
parameter $\widetilde\Omega_\m$ does not equal its bare (dynamical) value $\om$. This
opens a new avenue for testing such models against observations (see \cite{mimicry,om} in
this respect). Another important property of these models would be the growth of density
perturbations which is likely to differ from $\Lambda$CDM (see \cite{pert}, \cite{alam}
and references therein). This issue lies beyond the scope of the present paper but we may
return to it in the future. Braneworld models can be phantom-like and also exhibit
transient acceleration. Thus, brane phenomenology, with its basis in geometry, provides
an interesting alternative to `physical' dark energy scenario's such as quintessence.

The stability issues of the class of braneworld models without $Z_2$ mirror symmetry were
studied in \cite{CGP,KPS}. It is notable that ghost-free settings of the model with
transient acceleration (and phantom-acceleration) appear to exist \cite{CGP}. On the
other hand, the analysis of the recent paper \cite{KPS} reveals the presence of ghosts on
a background with a De~Sitter vacuum brane on the three branches $(++)$, $(+-)$, $(-+)$
(i.e., which have at least one `$+$', so that the bulk at least on one side of the brane
has `infinite volume' in terminology of \cite{KPS}). Whether this situation is critical
for the cosmology under investigation remains to be seen.  In this connection, it should
be noted that the $(++)$ branch, surviving in the $Z_2$ symmetric case, contains a ghost
and is, therefore, linearly unstable \cite{ghost}. On the other hand, the $(--)$ branch
is ghost-free in the $Z_2$ symmetric case and, apparently, also in the general case
without $Z_2$ symmetry. As we have demonstrated in this paper, this branch is responsible
for `phantom acceleration' ($w_0 < -1$); also see \cite{ss02,brane_obs,ls04}.

\acknowledgments

Yu.~S.\@ acknowledges support from the ``Cosmomicrophysics'' programme of the Physics and
Astronomy Division of the National Academy of Sciences of Ukraine. A.~S.\@ acknowledges
BIPAC and the support of the European Research and Training Network MRTPNCT-2006 035863-1
(UniverseNet).  A.~T.\@ acknowledges support from the RFBR grant 08-02-00923 and
scientific schools grant 4899.2008.2.

\appendix
\section*{Appendix}
\section{Visual representation of brane evolution}

The variables $\lbrace X,Y \rbrace$:
\begin{equation} \label{def-xy}
X \equiv {\rho_\tot \over 3 m^2} - H^2 \, , \qquad Y \equiv H^2 \, , \qquad
\rho_\tot = \rho + \sigma \, ,
\end{equation}
allow us to rewrite
equation (\ref{main}) in the form
\begin{equation} \label{main1}
X = - \sum_{i = 1,2} {\zeta_i \over \ell_i} \sqrt{Y + \lambda_i^{-2}} \, ,
\end{equation}
which has a convenient visual interpretation. Equation (\ref{main1}) describes four
branches in the physically restricted range $Y \ge 0$, with the symmetry of reflection
with respect to the $Y$ axis. If there exists a positive root $Y_c$ of the right-hand
side of (\ref{main1}) for $\zeta_1 \zeta_2 = -1$, then the two mixed branches intersect
each other at the point
\begin{equation} \label{intersect}
Y_c = H_c^2 = {\ell_1^2 \lambda_2^{-2} - \ell_2^2 \lambda_1^{-2} \over \ell_2^2 - \ell_1^2}
\, .
\end{equation}
The condition for the existence of this intersection point is that the constant on the
right-hand side of (\ref{intersect}) be positive.

The four branches (with and without intersection) are shown in Figs.\@
\ref{fig:intersect} and \ref{fig:nonintersect}, respectively.

\FIGURE{\label{fig:intersect}
\includegraphics[width=0.4\textwidth]{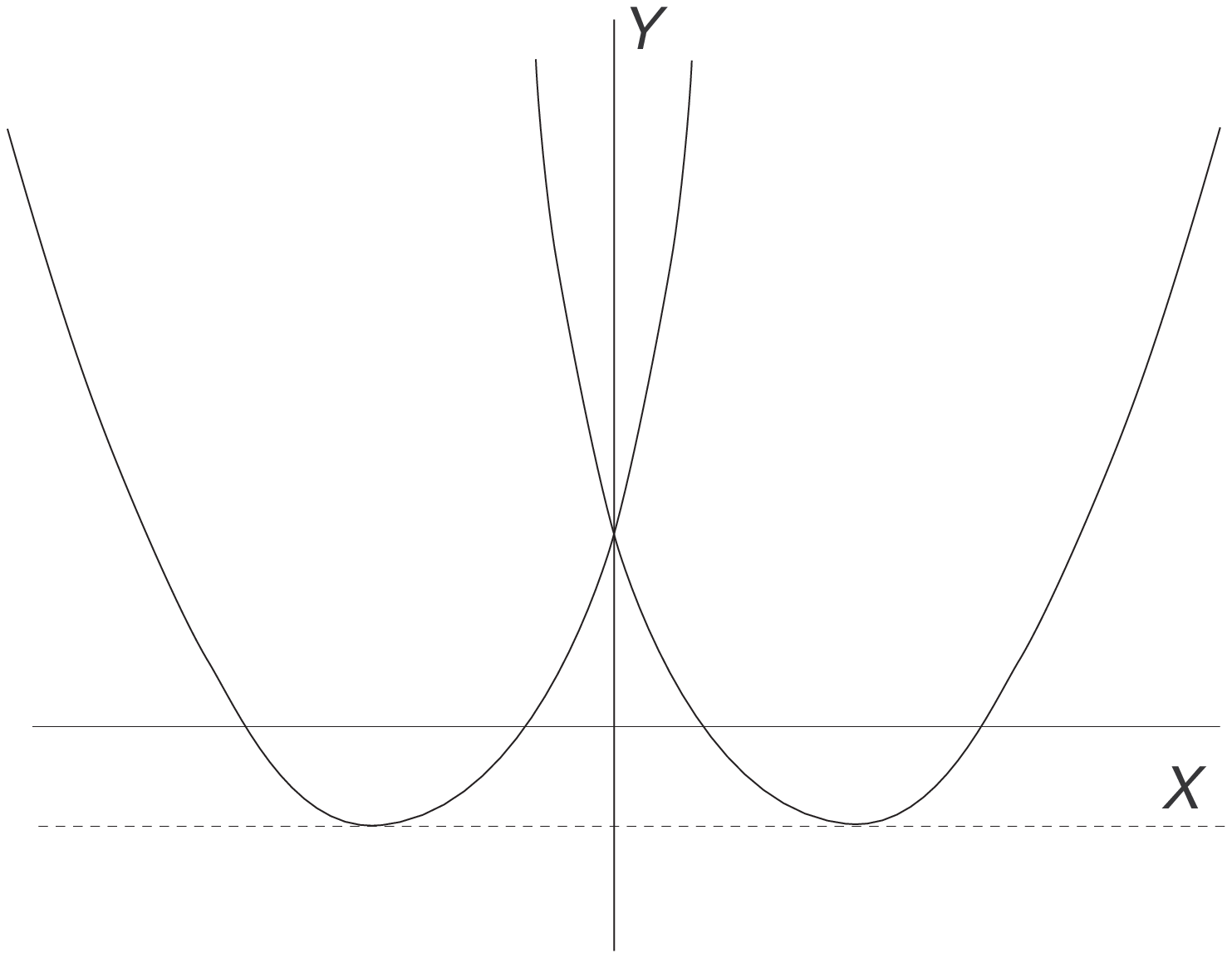} \hspace{1cm}
\includegraphics[width=0.45\textwidth]{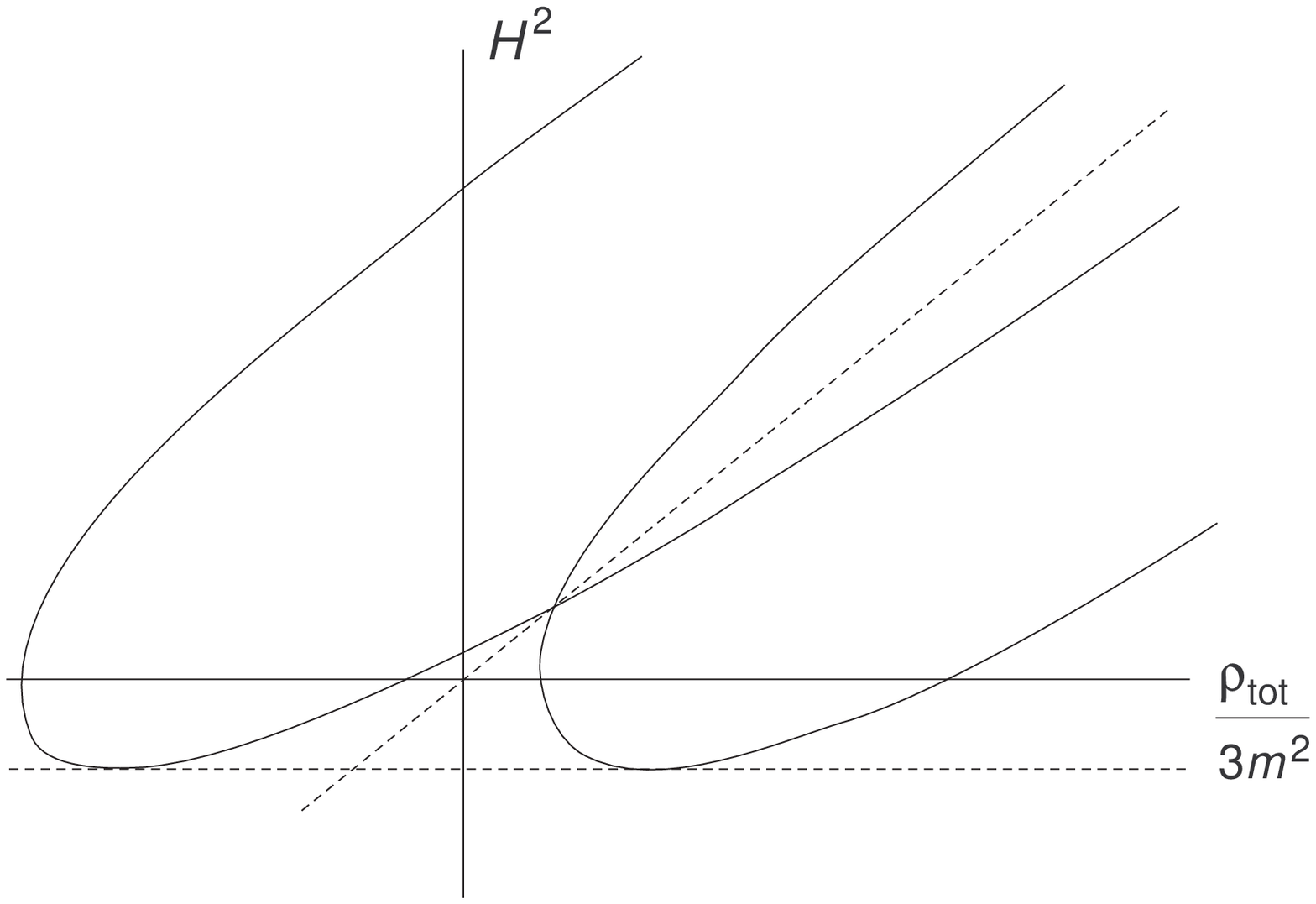}
\caption{Four branches described by Eq.~(\ref{main1}) in the ($X, Y$) plane and in the
($\rho_\tot, H^2$) plane in the case where two of them intersect.  The horizontal dotted
line indicates the position of the $H^2 = 0$ axis in the case $\Lambda_1 \Lambda_2 = 0$.
The region below this axis is nonphysical.}}

\FIGURE{\label{fig:nonintersect}
\includegraphics[width=0.45\textwidth]{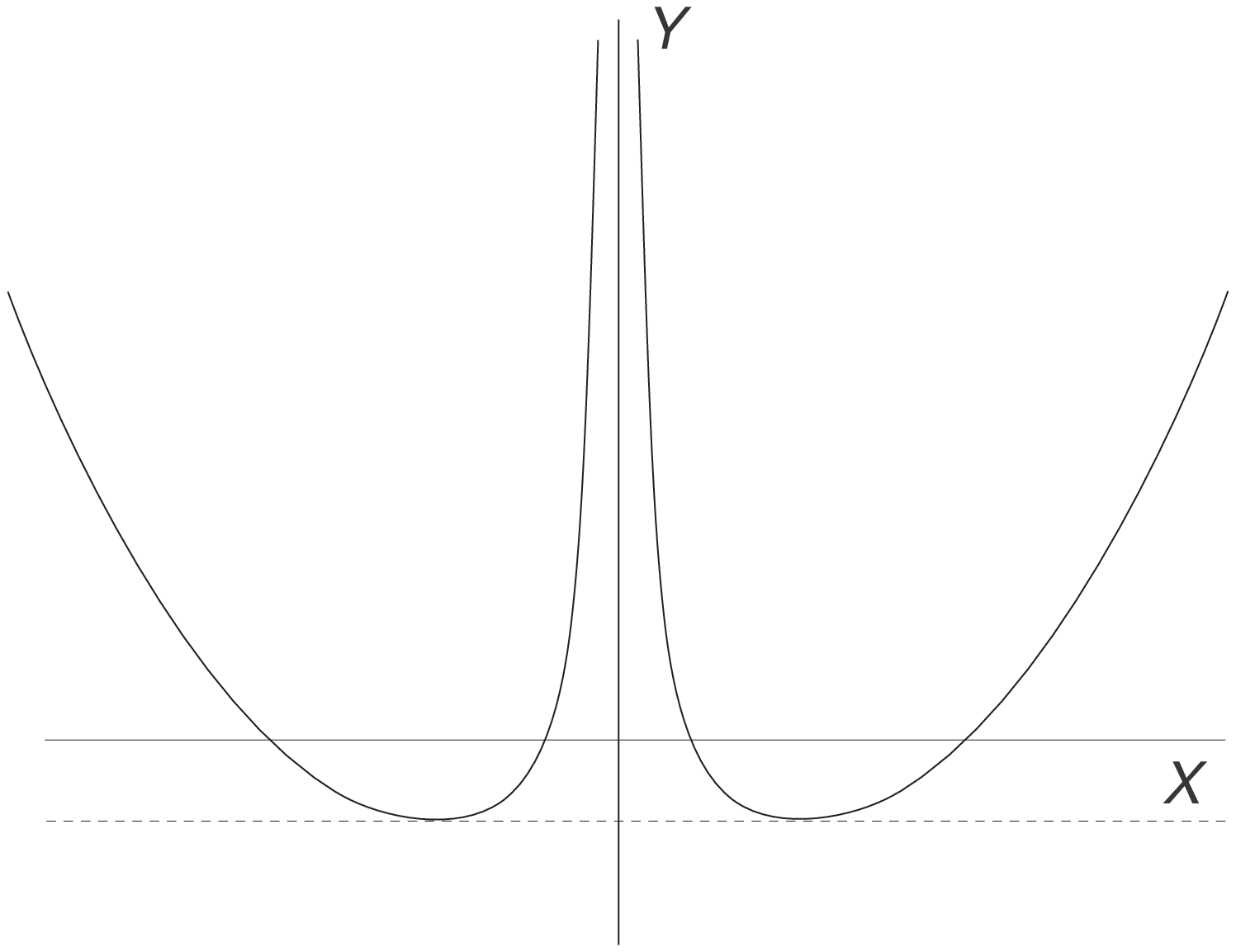} \hspace{1cm}
\includegraphics[width=0.45\textwidth]{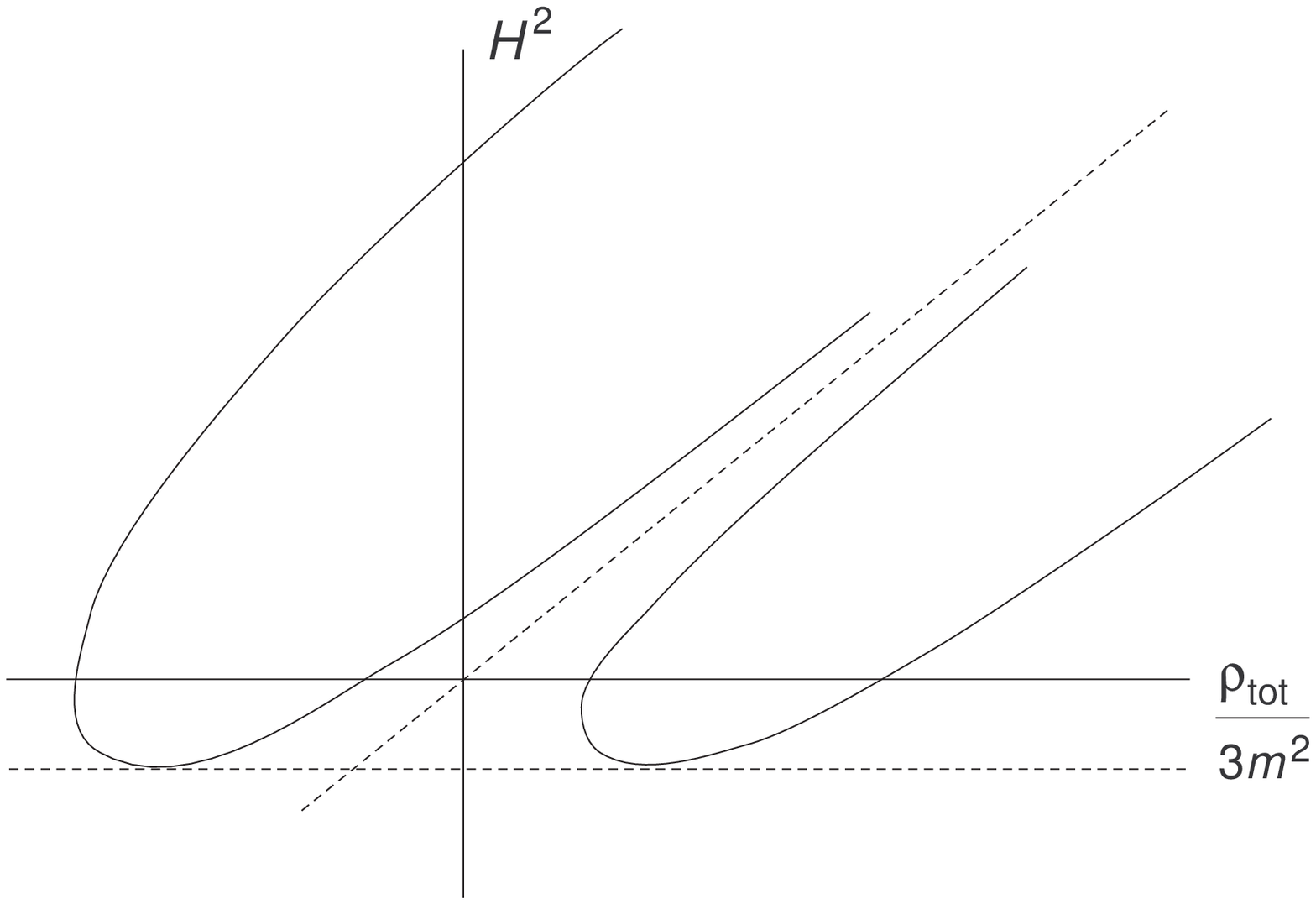}
\caption{Four branches described by Eq.~(\ref{main1}) in the ($X, Y$) plane and in the
($\rho_\tot, H^2$) plane in the case of absence of intersection. The horizontal dotted
line indicates the position of the $H^2 = 0$ axis in the case $\Lambda_1 \Lambda_2 = 0$.
The region below this axis is nonphysical.}}

The brane evolves along one of the four
branches towards decreasing values of $\rho_\tot$ (during expansion).
Depending (in particular) on the value of the brane tension $\sigma$,
three distinct possibilities can arise:
\begin{itemize}
\item[(i)] The trajectory may reach the value of $H = 0$, after which the universe
    recollapses and evolves along the same branch in the opposite direction.  This
    happens when the value of $\rho_{\rm tot}$ at this point is greater than its
    minimum value $\sigma$.

\item[(ii)] The trajectory may asymptotically tend to either de~Sitter space or the
    Minkowski universe with the minimum value $\rho_\tot = \sigma$.  The second
    possibility occurs when the minimum value of $\rho_\tot = \sigma$ is exactly the
    point where the corresponding graph crosses the axis $H^2 = 0$, which, therefore,
    requires some amount of fine tuning.  This possibility can be realised as {\em
    transient acceleration\/}.

\item[(iii)]  The trajectory may end in a quiescent singularity at a finite value of
    $H$. This happens when the {\em critical minimum point\/} of $\rho_{\rm tot}$ on
    the evolution curve is reached, and if this value of $\rho_{\rm tot}$ is greater
    than its minimum value $\sigma$.  The reasons for the existence of quiescent
    singularities can be seen from the right panels in Figs.\@ \ref{fig:intersect}
    and \ref{fig:nonintersect}.  They occur at the points of infinite derivative $d
    H^2 / d \rho_\tot$.
\end{itemize}

\end{document}